# Target Location by DNA-Binding Proteins: Effects of Roadblocks and DNA Looping

Gene-Wei Li[*], Otto G. Berg[†], and Johan Elf[‡§]

[*]*Department of Physics and Department of Chemistry and Chemical Biology, Harvard University, MA 02138 USA*
[†]*Department of Molecular Evolution, Uppsala University, Uppsala, Sweden*
[‡]*Department of Cell and Molecular Biology, Uppsala University, Uppsala, Sweden*
[§]*To whom correspondence should be addressed at: johan.elf@icm.uu.se*

The model of facilitated diffusion describes how DNA-binding proteins, such as transcription factors (TFs), find their chromosomal targets by combining 3D diffusion through the cytoplasm and 1D sliding along nonspecific DNA sequences. The redundant 1D diffusion near the specific binding site extends the target size and facilitates target location. While this model successfully predicts the kinetics measured in test tubes, it has not been extended to account for the highly crowded environment in living cells. Here, we investigate the effect of other DNA-binding proteins that partially occupy the bacterial chromosome. We show how they would slow down the search process, mainly through restricted sliding near the target. This implies that increasing the overall DNA-binding protein concentration would have a marginal effect in reducing the search time, because any additional proteins would restrict the search process for each other. While the presence of other proteins prevents sliding from the flanking DNA, DNA looping provides an alternative path to transfer from neighboring sites to the target efficiently. We propose that, when looping is faster than the initial search process, the auxiliary binding sites further extend the effective target region and therefore facilitate the target location.

## Introduction

DNA-binding proteins control both when and how the genome functions. Moreover, they limit how fast the genomic information can be accessed. For instance, the rate at which a transcription factor (TF) finds its specific binding site determines how rapidly the corresponding gene can be turned on or off. When this rate is low compared to the speed of the regulated process, the binding and unbinding kinetics of the protein may be more important for the overall system behavior than its often well-characterized equilibrium properties.

Although seemingly a simple process, target location by DNA-binding proteins requires scanning through millions to billions of similar nonspecific DNA sequences. It was hypothesized [1, 2] and later demonstrated [3] that DNA-binding proteins partly diffuse (slide) along the nonspecific DNA while scanning for the specific site (Fig. 1). The process of alternating 3D diffusion in cytoplasm and 1D diffusion along DNA is called facilitated target location [4]. Originally discovered *in vitro*, this mechanism facilitates the process of target location by extending the target size via efficient sliding from neighboring nonspecific DNA, while taking advantage of three-dimensional diffusion to carry the molecule over large distances. The complexity of this search process has attracted both theoretical and experimental attention over the past few decades [2, 5-14].

In a living cell, however, the target location is further complicated by the vast amount of other proteins, both cytoplasmic and DNA-bound. More than 50% of the dry weight of an E. coli cell is protein [15]. The 3D diffusion in the cytoplasm is slowed down by at least one order of magnitude compared to that in water. Furthermore, more than 100,000 proteins are associated with DNA per cell, mostly bound nonspecifically [16, 17]. As a result, the bacterial chromosome is at least 20-30% occupied, with mostly uniformly distributed short naked DNA fragments [18]. The sliding motion along DNA must be hindered as well. Therefore, for facilitated target location, it is important to consider the effects of these macromolecules that are absent in the *in vitro* settings.

In the current paper we adapt the theory for facilitated target location to the *in vivo* situation, with emphasis on the crowding on chromosomal DNA in bacterial cells. First, the high occupancy by other proteins results in roadblocks which restrict the open sliding on DNA. The effect is analyzed for both stationary roadblocks and sliding roadblocks. Second, the other proteins compete for both nonspecific sites and specific site with the target locator. We combine these effects and derive analytical expressions for how the rates of binding to a specific site are reduced by roadblocks. We also propose that DNA looping from auxiliary sites can bypass the roadblocks and speed up the search process. Specific examples will be given with the *E. coli* lactose utilization (*lac*) operon, the paradigm of gene regulation. We focus on the *lac* repressor (LacI), which regulates the expression of the *lac* operon in response to the intracellular lactose concentration. The results are compared to the recent single molecule measurements on search kinetics *in vivo*.

## Theory

We use the Smoluchowski bi-molecular association rate constant as a basis, and incorporate various effects of sliding and roadblocks. The second-order association rate constant for a diffusing molecule and a stationary target is proportional to both the diffusion constant, $D$, and the target size, $L$ [19]

$$k_a \propto DL \qquad (1)$$

The two relevant parameters, $D$ and $L$, reflect the global and local properties of the system, respectively. The diffusion constant addresses how fast the molecule can approach the target globally, while the target size is a measure of how easily the reaction can take place locally.

Under the framework with global and local contributions, it is possible to consider the effects of nonspecific DNA binding and sliding. Nonspecific binding slows down the global diffusion towards the target. Because the one-dimensional diffusion of sliding is typically two orders of magnitude smaller than the three-dimensional diffusion in cytoplasm [20, 21], the TF only diffuses significantly while in cytoplasm. This can be considered as a reduced effective diffusion constant when the nonspecific DNA is uniformly distributed over the length scale it takes to find the target. The effective diffusion constant is the 3D diffusion constant weighted by the fraction of time, $w$, that the TF spends in cytoplasm [13]:

$$D \approx wD_3 \approx \frac{D_3}{1 + K_{RD}c_{ns}} \qquad (2)$$

In the last expression, we re-write the fraction of time spent in 3D diffusion using the equilibrium binding constant between TF and nonspecific site, $K_{RD}$, and the concentration of nonspecific sites, $c_{ns}$. The concentration of nonspecific sites in the *E. coli* cell is on the order of 10 mM, and the nonspecific binding constant for the *lac* repressor is estimated to be 1 mM$^{-1}$ [6, 13]. This results in a ten-fold decrease for diffusion *in vivo*.

Close to the binding site, sliding increases the target size, $L$. Because of the redundant nature of one dimensional diffusion, TFs reaching a nearby nonspecific site will have high probability of reaching the specific site. As a result, the flanking DNA adjacent to the binding site acts as a strong sink for the TF. The size of the sink is roughly given by the average sliding length along DNA, $\sqrt{D_1/\Lambda}$, where $D_1$ is the 1D diffusion constant (in μm$^2$/sec), and $\Lambda$ is the macroscopic dissociation rate between TF and nonspecific sites (in 1/sec). More detailed calculations [2] showed that the effective target size is given by

$$L = l + 2\sqrt{D_1/\Lambda} \equiv l + l_{sl}, \qquad (3)$$

where $l$ is the size (in μm) of one basepair. In the derivation of Eq.(3), it is assumed that the effective target is positioned as a straight line in space. In other words, the sliding length is assumed to be shorter than the persistence length of DNA. *In vivo*, the macroscopic dissociation rate for the *lac* repressor is faster than (5ms)$^{-1}$, whereas the 1D diffusion constant *in vitro* is 0.046μm$^2$/s [13], which could be even smaller *in vivo* considering the difference in viscosity. As a result the sliding length of a *lac* repressor is shorter than 85 basepairs, which should be compared with the persistence length of 150 basepairs [22].

We have introduced the macroscopic dissociation rate, $\Lambda$. This is in contrast to the microscopic dissociation that brings the two molecules (protein and DNA) apart by an interaction distance, $b$ [2]. Since a microscopic dissociation is often followed by rapid re-binding to the same site or a neighboring site, it takes many microscopic dissociation events before the two molecules lose their spatial correlation, or dissociate macroscopically. In the case of a TF dissociating from chromosomal DNA in a cell, one can consider the correlation being lost as soon as they are separated by half of the average spacing between DNA chains (Fig. 1). We define this characteristic length scale, $R_c$, as $\pi R_c^2 Ml = $ nucleoid volume, where $M$ is the total number of basepairs and $l$ is the length per basepair [2]. For an *E coli* chromosome with 1μm$^3$ of nucleoid volume, $R_c$=14nm. The interaction distance, $b$, is considered to be the DNA radius, or 1nm. With this definition, the diffusion-limited macroscopic association rate constant to nonspecific sites is [2]

$$k_a^{ns} = \frac{2\pi D_3 l}{\ln(R_c/b)}. \qquad (4)$$

The macroscopic dissociation rate can be obtained using the equilibrium binding constant

$$\Lambda = k_a^{ns} / K_{RD} = \frac{2\pi D_3 l}{\ln(R_c/b)} / K_{RD}. \qquad (5)$$

For a TF of a typical size in *E. coli*, the diffusion constant in cytoplasm is ~3μm²/s [13]. Therefore, the diffusion-limited nonspecific dissociation rate is $\Lambda = (0.7 ms)^{-1}$ for $K_{RD}$=1mM⁻¹.

The expression for the macroscopic nonspecific association rate constant (Eq. (4)) also provides the proportionality constant in Eq. (1)

$$k_a = \frac{2\pi}{\ln(R_c/b)} DL. \qquad (6)$$

Combining Eq. (2) and Eq. (3), one can obtain the expression for the specific association rate constant of facilitated target location.

$$k_a = \frac{2\pi}{\ln(R_c/b)} \frac{D_3}{1 + K_{RD} c_{ns}} (l + l_{sl}) \qquad (7)$$

This expression was first obtained by Berg, Winter and von Hippel by solving diffusion equations with proper boundary conditions [2]. Over the past ten years, it has been derived again by many other methods [8-12]. Using the equilibrium constants, we can also obtain the specific dissociation rate constant

$$k_d = k_a \frac{1 + c_{ns} K_{RD}}{K_{RO}} = \frac{2\pi}{\ln(R_c/b)} \frac{D_3 (l + l_{sl})}{K_{RO}}, \qquad (8)$$

where $K_{RO}$ is the equilibrium binding constant of the specific site, and $c_{ns}$ is the concentration of nonspecific sites.

## Result

### Semi-stationary roadblock

Other DNA-binding proteins near the target site may serve as roadblocks for 1D sliding along DNA. Because roadblocks prevent diffusing into the target via nonspecific sites, the effective target size is reduced. This will make both association to and dissociation from the target site slower (Fig. 2A). For simplicity, we assume that on average there is no interaction between the diffusing molecule and the roadblocks, except that the roadblock can be considered as a reflective boundary for 1D diffusion. For a roadblock at a fixed distance from the target, consider the geometry illustrated in Figure 2B: the spacing between the boundary of the roadblock and the boundary of the target is defined as *s*. This is the distance that the DNA-binding protein can slide without being absorbed at x=0 or being reflected at x=s. In the diffusion-limited case, as soon as the protein reaches x=0, specific binding is established and further sliding is abolished. Hence we can regard sliding from either side of the target as independent and treat only the boundary conditions described above. For a specific site located symmetrically between two roadblocks, the effective target size of a binding was first derived by Berg *et al.*: [2]

$$L(s) = l + l_{sl} \tanh\left(\frac{2s}{l_{sl}}\right) \qquad (9)$$

As expected, the target size becomes the length of the flanking DNA (*2s*) if the spacing between the roadblocks is much shorter than the free sliding length ($l_{sl}$). In the other limit, the size equals the sliding length at large spacing (*2s*>> $l_{sl}$).

On a bacterial chromosome, the roadblocks are not always situated at a fixed distance from the targets. Instead, there is a distribution of the spacing (*s*), depending on what fraction of DNA is occupied by nonspecific proteins. The effective target size is therefore an average over the different spacings. We assume that the majority of proteins are randomly distributed throughout the genome, as indicated by measurements on major nucleoid proteins [18]. In this section, we consider the case that such spatial distribution is stationary at the time scale of a single sliding event (~ms), but re-randomized at a time scale of diffusion across the nucleoid (~sec). Define the vacancy, *v*, as the fraction of roadblock-free DNA, and the footprint, *d*, as the average footprint of all DNA-binding proteins, as well as the length of the specific site. In the followings we will derive the distribution of flanking DNA lengths as a function of the vacancy.

When roadblocks are randomly positioned on DNA, the average gap size between them is $dv/(1-v)$. The distribution of gap sizes can be obtained by counting the number of ways to position all the roadblocks on DNA [23, 24]. In the Supporting Information we show that the gap sizes are approximately exponentially distributed

$$p_{gap}(s) \approx \frac{1-v}{dv}\exp\left(-\frac{s}{dv/(1-v)}\right). \tag{10}$$

Instead of the distribution of all the gaps, we are interested in the distribution of flanking lengths at a particular location on DNA in order to calculate the effective target size. In fact, the probability of having any gaps at all at a particular location is equal to the vacancy $v$. When a gap opens up, the distribution of spacing to the next roadblock is then $p_{gap}(s)$. Therefore, the length distribution for the open DNA flanking a specific position is the product of the two probabilities, $vp_{gap}(s)$ (Supporting information).

It is now possible to calculate the average effective target size at a given vacancy $v$, by averaging $L(s)$ weighted by the distribution of $s$. As illustrated in Fig. 2B, for a protein to slide into the target, the roadblock-free region has to extend from $x=-d$ to $x=s$. In other words, for a flanking length of $s$ from roadblock to target, the total length of "open" DNA has to be $s+d$. Therefore, we arrive at the average effective target size at a given vacancy $v$:

$$\langle L(v)\rangle_s = \int_0^\infty vp_{gap}(s+d)L(s)ds = ve^{1-\frac{1}{v}}\left[l+l_{sl}\left(1+2\sum_{k=1}^\infty (-1)^k\left(1+4\frac{d}{l_{sl}}\frac{v}{(1-v)}k\right)^{-1}\right)\right]. \tag{11}$$

The subscript $s$ denotes the case with semi-stationary roadblock. This is plotted in Figure 3A. The effective target size decreases as the vacancy decreases. When the average gap size is much smaller than the free sliding length, $l_{sl}$, the average effective target size becomes:

$$\langle L(v)\rangle_s \xrightarrow{l_{sl}\gg\frac{dv}{1-v}} ve^{1-\frac{1}{v}}\left(l+2\frac{dv}{1-v}\right) \tag{12}$$

This expression can be understood from the following. The multiplicative factor, $ve^{1-\frac{1}{v}}$, is the probability that the entire target site (of length $d$) is open $\left(\int_d^\infty vp_{gap}(s)ds\right)$. When the target site is open, the distance to the next roadblock is distributed according to $p_{gap}(s)$ with an average of $\frac{dv}{1-v}$. Since the protein can slide from either side of the target, the extended target size is twice the average distance.

Besides affecting the target size, the nonspecific binding proteins also change the effective diffusion constant for the target search. The number of accessible nonspecific DNA for the protein of footprint $d$ is reduced as the vacancy decreases. Similar to the probability that the entire target site is open, the fraction of accessible DNA goes down as $ve^{1-\frac{1}{v}}$. Therefore, the effective concentration of DNA ($c_{ns}$) in calculating the effective diffusion constant is modified by the same factor.

$$D = \frac{D_3}{1+K_{RD}c_{ns}ve^{1-1/v}}. \tag{13}$$

Combining the changes in diffusion and target size, we obtain the following expression for the association rate constant.

$$k_a = \frac{2\pi}{\ln(R_c/b)}\frac{D_3}{1+K_{RD}c_{ns}ve^{1-1/v}}\langle L(v)\rangle_s \tag{14}$$

The search time, which is defined as the average time for a single DNA-binding protein to find a single specific site in one chromosome, is plotted in Figure 3B. It is worth noting that when nonspecific binding is dominant ($K_{RD}c_{ns}ve^{1-1/v}\gg 1$), the factor $ve^{1-\frac{1}{v}}$ in the denominator cancels with that in $\langle L(v)\rangle_s$. In other words, the enhancement in global diffusion cancels out the effect of partially blocked targets. This leaves only the vacancy dependence in the restricted sliding length, or the spacing between the target and the nearest roadblock.

**Dynamic roadblock**

Next we consider the case in which the roadblocks are also diffusing on DNA. Contrary to the previous case with stationary roadblocks, there is now a finite probability of reaching a distant target because the roadblocks can move away. This will result in a larger extended target size if the searcher stays long enough. We assume that the roadblocks and target-locator have the same 1D diffusion constant and ignore the binding and unbinding of the roadblocks during each sliding event of the target-locator. Because the majority of roadblocks (nucleoid proteins) have higher nonspecific affinity than transcription factors (such as LacI) in general [25-28], we expect slower exchange between 3D and 1D motions. In this case, the effective target size can be obtained by analyzing how far the target-locator moves before dissociation on a track with diffusing non-passing roadblocks.

This system with identical, non-passing diffusing particles in an 1D channel is known as Single-File Diffusion [29, 30], and has been shown to exhibit subdiffusion at long times. For time longer than the diffusion time across the average gap size, the mean-square displacement scales as the time to the one-half power: $<x^2(t)> = 2Ft^{1/2}$, where $F = dv(1-v)^{-1}\sqrt{D_1/\pi}$. Since we are interested in the scaling for all vacancy levels, a complete knowledge of $<x^2(t)>$ across all time scales is needed. Although a theoretical description is not yet attainable, a phenomenological approximation has been shown to be valid on all relevant time scales [31]. Lin *et al.* have demonstrated experimentally that the single-particle Green's function retains a Gaussian shape at all times, with the variance well described by:

$$<x^2(t)> = \frac{2D_1 t}{1+(t/t_x)^{1/2}}, \quad (15)$$

where $t_x = (F/D_1)^2 = (dv/(1-v))^2/\pi D_1$ is the diffusion time across the average gap size [31]. Let us define the residence time $t_R$ ($=1/\Lambda$) as the average time per visit on nonspecific DNA. The average extended target size can be approximated as (Supporting Information):

$$\langle L(v) \rangle_d = ve^{1-\frac{1}{v}}\left(l + 2\sqrt{\frac{D_1 t_R}{1+(t_R/t_x)^{1/2}}}\right), \quad (16)$$

where $ve^{1-\frac{1}{v}}$ is the probability that the protein will bind to the nonspecific sites in the first place (opening with at least length $d$). $\langle L(v) \rangle_d$ is plotted in Figure 3A as a function of the vacancy $v$. Although one would expect the effective target to be bigger than that in the stationary case when gaps are smaller (low vacancy), the difference between dynamic and stationary roadblocks is actually negligible. This is because the opening of free DNA longer than the protein footprint is the rare event at low vacancy, and hence the dominant factor in both cases.

Combining with the change in diffusion constant, the association rate constant with dynamic roadblocks is

$$k_a = \frac{2\pi}{\ln(R_c/b)} \frac{D_3}{1+K_{RD}c_{ns}ve^{1-1/v}} \langle L(v) \rangle_d. \quad (17)$$

Unlike effective target size, the major vacancy dependence here is in the diffusion length restricted by roadblocks, for the same reason in the stationary case where $ve^{1-\frac{1}{v}}$ is canceled out. The possibility of traveling long distances with diffusing roadblocks therefore leads to shorter search time at low vacancy compared to the stationary case (Fig. 3B).

## Discussion

**Enhancement of search kinetics by sliding**

We have seen that the search time is increased with the partially-covered chromosome. But does sliding still facilitate the target location? To answer this question, we consider two hypothetical scenarios in which there is no sliding ($D_1 = 0$) or no nonspecific binding ($K_{RD} = 0$ and $D_1 = 0$), respectively. For a fair comparison, we further assume that the hypothetical specific target sizes (*l*) are the same in all cases. As seen in Fig. 4, sliding helps reduce the search time for almost all vacancy levels. It results in faster target location when compared to only 3D diffusion (*green*) or no sliding (*yellow*).

**Search kinetics with increasing TF concentration**

Search times in the order of minutes may seem long considering the short generation times of bacterial cells. It should however be remembered that these times are defined for one molecule searching for one target. A faster response of gene regulation can be achieved by increasing the number of TFs that search for the target. By doubling the copy number of a single TF species, the time for binding the target will be halved. However, due to crowding on the chromosome, a global increase in all TF concentrations can lead to increased search times due to reduced vacancy, if vacancy is low. Therefore, only a selected group of DNA-binding proteins can be expressed at high copy numbers, whereas the majority should have as low concentrations as possible. The selection for expression levels of different DNA-binding proteins are known to be under a variety of pressures.

Among those are the requirements of robustness against fluctuations [32], dynamic range in response [33], cost of synthesis [34] and many others. The compromise between faster kinetics and chromosome vacancy adds another dimension to the evolution of the proteome composition.

Let us examine the effect of a global change in the total number of DNA-binding proteins. Assuming that the relative protein concentrations are selected by the requirements listed above, the overall concentration of them has to be low enough such that the chromosome remains largely unoccupied, but also high enough such that their kinetics remains fast individually. Figure 5 shows how fast the binding to a single operator can be reached as a function of total number of DNA-binding proteins. It is assumed that the TF concentration is at a small but fixed fraction of the total DNA-binding protein concentration. The result shows that a minimum time of binding is achieved at $10^4$-$10^5$ total proteins per chromosome equivalent. The actual number of these proteins in *E. coli* was reported to be about 30,000 per chromosome equivalent in log phase [16, 17]. It is thus evident from the plot that further increasing the overall concentration from the current value has little effect in shortening the reaction time.

**Facilitation by auxiliary sites and rapid DNA looping**

An alternative strategy for facilitating the search kinetics is to increase the number of binding sites to search for. For instance, two specific sites spaced apart by longer than one sliding length would appear as two independent targets. The time to find any one of them is thus halved (Fig 6A). However, often only one binding site is responsible for the desired biological functions, such as blocking the promoter sequence. We define this as the regulatory site. The search time for the regulatory site can only be reduced if binding at the auxiliary sites promotes binding to the regulatory site, such that the inter-operator transfer on average is much faster than the initial search process. A simple way for the TF to obtain this effect is to have two separate DNA-binding domains so that it can reach the regulatory site while still sitting on the auxiliary site. Since the binding sites must be more than one sliding distance apart ($>l_{sl}$), this would lead to the formation of a DNA loop. Using multiple auxiliary sites, such as the dual auxiliary operators in the case of *lac*, may speed up the search process further. Too many auxiliary sites would however slow down the search process due to competition with the regulatory site.

The role of auxiliary sites in facilitating target location is different from inter-segment transfers. This term refers to a hypothetical mechanism where the protein binds to two nonspecific DNA segments and transfers from one to the other without intermediate dissociation into cytoplasm. Although potentially it could also allow the protein to bypass roadblocks and transfer to the regulatory site, the transfer efficiency from nonspecific sites is low compared to that from auxiliary operator sites. With higher affinities than nonspecific sites, the auxiliary sites retain proteins for longer periods of time. The longer interaction time, especially when longer than the loop formation time, provides a higher probability of reaching the regulatory site upon binding. Therefore, DNA looping may be a fundamentally different way of achieving faster target location, in addition to the commonly known mechanisms of sliding and inter-segment transfer. We identify this property as another benefit of DNA looping, which has been shown to be capable of reducing both sensitivity to TF concentration and fluctuation in gene expression [35].

**Comparison of search time and loop formation time**

A necessary condition for facilitating the search process by auxiliary sites is that the loop formation rate has to be fast compared to the time it takes to find the first site. Compared to the initial search, the space to search for the second site, after binding to the first site, is much smaller because of the proximity of the two binding sites [36]. However, these rates can not be directly compared by their respective reaction volumes due to the different nature of the underlying processes. While the initial binding event involves proteins diffusing through cytoplasm and along DNA, the loop formation relies on fluctuations of the linker DNA conformation. Moreover, sliding between the protein and the second site is geometrically hindered because binding at the first site locks the relative position between them. Sliding into the second site would therefore require both twisting and lateral diffusion of the DNA (Fig. 6B). Because the *in vivo* chromosome dynamics at the time scale of looping are largely unknown, it is not yet possible to calculate exactly how much the loop formation kinetics are slowed down by these new geometrical constraints. However, analyses and simulations based on DNA properties *in vitro* have suggested that the time scale of loop formation is shorter than seconds [37-39].

In comparison, the search time *in vivo* was recently measured with live-cell single molecule imaging techniques [13]. With fluorescently labeled *lac* repressor in *E. coli*, the search time was probed by removing the inducers from the growth media and watching how soon the rebinding occurs. This experiment showed that it takes at most one minute for three dimeric *lac* repressors to find any of the two strong operators. The search time for a single repressor to find a single operator site is therefore less than 6 minutes, which can be compared to the search times plotted in Figure 3B. The use of dimeric *lac* repressors eliminates the possibility of intersegment transfer and DNA looping in the search process. The 1D and 3D diffusion constants of the wild type tetrameric repressor should however be similar to those of the dimeric repressors labeled with two fluorescent proteins. In the wild type *E. coli*, there are at least five tetrameric *lac* repressors [40]. The time to reach a single operator is therefore less than 70 seconds. Since the initial binding is slower than the loop formation by approximately two orders of magnitude, we expect that the search process for the lac repressor is indeed facilitated by DNA looping.

In passing we also note that if sliding into the looped conformation is restricted by the geometrical constraints described previously, this would also be the case for loop deformation, as a consequence of detailed balance. Instead of first sliding into nonspecific sites before fully dissociating, the DNA-binding domain has to break both the specific interaction (mostly hydrogen bonds) and the nonspecific interaction (electrostatic interaction) (Fig. 6B). If the cost of energy for DNA looping is negligible, it would therefore take longer time to dissociate from one of the binding sites in the looped conformation than from the same binding site in a non-looped conformation. However, in the case of the *lac* operon where the cost of energy for looping can not be neglected [33, 41], the rate of dissociation from looped may not be reduced as much.

## Conclusion

In conclusion, we have extended the facilitated diffusion model by Berg *et al.* [2] to include the effect of high occupancy by proteins on the bacterial chromosome. We find that the high occupancy on DNA (i) reduces time spent on nonspecific DNA by limiting the number of accessible nonspecific sites, (ii) reduces the sliding distances of TF, and (iii) partially blocks the specific binding site. We have derived analytical expressions for how these effects change the search time for a specific site, and most importantly, how the average sliding distances depend on the DNA occupancy in the case of stationary or diffusing roadblocks. As a consequence, too many DNA-binding proteins on a chromosome may lead to reduced rate of target binding when the vacancy is too low. We showed that the total number of all DNA-binding proteins in *E. coli* is close to the limit, from which further increasing the number is catastrophic. We also introduced auxiliary sites as a way to bypass the sliding roadblocks. By having auxiliary sites positioned more than the average sliding distance apart from the regulatory site, all sites are recognized as independent search targets for the DNA-binding protein. If inter-site transfer of the protein from an auxiliary site to the regulatory site is fast compared to the initial search processes, the overall rate for reaching the regulatory site is increased. DNA looping between high-affinity sites can therefore further facilitate target location when sliding is hindered by roadblocks. Finally we note that the need for fast target location could be one of the reasons that TF-mediated DNA looping is commonly observed in bacteria.

**Acknowledgments**
We are grateful for helpful discussions and comments from Paul Choi, Sunney Xie, Jose Vilar, Måns Ehrenberg, Brian English, Ji Yu and Peter Wolynes. This work was supported by ERC-SIRG, the Swedish Foundation for Strategic Research, the Wallenberg Foundation, and NSF Graduate Research Fellowship (G.-W. L.).

# References

1. Richter, P.H., and Eigen, M. (1974). Biophysical Chemistry *2*, 255-263.
2. Berg, O.G., Winter, R.B., and von Hippel, P.H. (1981). Biochemistry *20*, 6929-6948.
3. Blainey, P.C., van Oijen, A.M., Banerjee, A., Verdine, G.L., and Xie, X.S. (2006). Proceedings of the National Academy of Sciences of the United States of America *103*, 5752-5757.
4. von Hippel, P.H., and Berg, O.G. (1989). The Journal of Biological Chemistry *264*, 675-678.
5. Winter, R.B., and von Hippel, P.H. (1981). Biochemistry *20*, 6948-6960.
6. Winter, R.B., Berg, O.G., and von Hippel, P.H. (1981). Biochemistry *20*, 6961-6977.
7. Barkley, M.D. (1981). Biochemistry *20*, 3833-3842.
8. Slutsky, M., and Mirny, L.A. (2004). Biophysical Journal *87*, 4021-4035.
9. Halford, S.E., and Marko, J.F. (2004). Nucleic Acids Research *32*, 3040-3052.
10. Zhou, H.X., and Szabo, A. (2004). Physical Review Letters *93*, 178101.
11. Klenin, K.V., Merlitz, H., Langowski, J., and Wu, C.X. (2006). Physical Review Letters *96*, 018104.
12. Hu, T., Grosberg, A.Y., and Shklovskii, B.I. (2006). Biophysical Journal *90*, 2731-2744.
13. Elf, J., Li, G.W., and Xie, X.S. (2007). Science *316*, 1191-1194.
14. Eliazar, I., Koren, T., and Klafter, J. (2007). J Phys: Condens. Matter *19*.
15. Bremer, H., and Dennis, P.P. (1996). Modulation of Chemical Composition and Other Parameters of the Cell by Growth Rate. In Escherichia coli amd Salmonella Cellular and Molecular Biology, F.C. Neidhardt, ed. (American Society Microbiology).
16. Ali Azam, T., Iwata, A., Nishimura, A., Ueda, S., and Ishihama, A. (1999). Journal of Bacteriology *181*, 6361-6370.
17. Johnson, R.C., Johnson, L.M., Schmidt, J.W., and Gardner, J.F. (2005). Major Nucleoid Proteins in the Structure and Function of the Escherichia coli Chromosome. In The Bacterial Chromosome, N.P. Higgins, ed. (Washington, D.C.: ASM Press).
18. Varshavsky, A.J., Nedospasov, S.A., Bakayev, V.V., Bakayeva, T.G., and Georgiev, G.P. (1977). Nucleic Acids Research *4*, 2725-2745.
19. Berg, O.G., and von Hippel, P.H. (1985). Annual Review of Biophysics and Biophysical Chemistry *14*, 131-160.
20. Schurr, J.M. (1979). Biophysical Chemistry *9*, 413-414.
21. Bagchi, B., Blainey, P.C., and Xie, X.S. (2008). J Phys Chem B *112*, 6282-6284.
22. Calladine, C., Drew, H., Luisi, B., and Travers, H. (2004). Understanding DNA, (Elsevier).
23. McGhee, J.D., and von Hippel, P.H. (1974). Journal of Molecular Biology *86*, 469-489.
24. Flyvbjerg, H., Keatch, S.A., and Dryden, D.T. (2006). Nucleic Acids Research *34*, 2550-2557.
25. Pinson, V., Takahashi, M., and Rouviere-Yaniv, J. (1999). Journal of Molecular Biology *287*, 485-497.
26. Betermier, M., Galas, D.J., and Chandler, M. (1994). Biochimie *76*, 958-967.
27. Friedrich, K., Gualerzi, C.O., Lammi, M., Losso, M.A., and Pon, C.L. (1988). FEBS Lett *229*, 197-202.
28. Yang, S.W., and Nash, H.A. (1995). Comparison of protein binding to DNA in vivo and in vitro: defining an effective intracellular target. The EMBO Journal *14*, 6292-6300.
29. Harris, T.E. (1965). Journal of Applied Probability *2*.
30. Wei, Q., Bechinger, C., and Leiderer, P. (2000). Science *287*, 625-627.
31. Lin, B., Meron, M., Cui, B., Rice, S.A., and Diamant, H. (2005). Physical Review Letters *94*, 216001.
32. Batchelor, E., and Goulian, M. (2003). Proceedings of the National Academy of Sciences of the United States of America *100*, 691-696.
33. Bintu, L., Buchler, N.E., Garcia, H.G., Gerland, U., Hwa, T., Kondev, J., and Phillips, R. (2005). Curr Opin Genet Dev *15*, 116-124.
34. Ehrenberg, M., and Kurland, C.G. (1984). Costs Q Rev Biophys *17*, 45-82.
35. Vilar, J.M., and Leibler, S. (2003). Journal of Molecular Biology *331*, 981-989.
36. Muller-Hill, B. (1998). Molecular Microbiology *29*, 13-18.
37. Merlitz, H., Rippe, K., Klenin, K.V., and Langowski, J. (1998). Biophysical Journal *74*, 773-779.
38. Jun, S., Bechhoefer, J., and Ha, B.-Y. (2003). Europhys Lett *64*.
39. Hyeon, C., and Thirumalai, D. (2006). J Chem Phys *124*, 104905.
40. Gilbert, W., and Muller-Hill, B. (1966). Proceedings of the National Academy of Sciences of the United States of America *56*, 1891-1898.
41. Saiz, L., Rubi, J.M., and Vilar, J.M. (2005). Proceedings of the National Academy of Sciences of the United States of America *102*, 17642-17645.

# Figure Legends

**Fig. 1.** Original model for facilitated target location. In the model investigated by Berg *et al.* [2], the DNA-binding protein searches for its binding site (BS) by combining 1D diffusion ($D_1$) along nonspecific DNA and 3D diffusion ($D_3$) in cytoplasm. The macroscopic dissociation rate from nonspecific sites ($\Lambda$) is defined as the rate of dissociation events that are followed by diffusion away from DNA at distance $R_c$, which is half of the average spacing between neighboring DNA strands. In a living cell, with $10^6$-$10^9$ basepairs of DNA, the protein needs to go through many rounds of 1D and 3D searches before locating the target.

**Fig. 2.** Reduction of effective target size by roadblocks at a fixed distance from the binding site. (A) The effective target size is the size of the flanking DNA from which the DNA-binding protein can slide into the specific site before dissociation. For roadblocks positioned farther than one sliding length, $l_{sl}$, the effective target size is unchanged (*above*). For roadblocks positioned close to the target size, the length of flanking DNA that the protein can diffuse from is significantly reduced (*below*) (B) Consider a specific site spanning from x=-*d* to x=0. Here we only discuss sliding from the right-hand side of the target. Sliding from the other side can be treated symmetrically. The distance that a target-locator can diffuse without been absorbed or reflected is defined as *s*. In a cell, there is a distribution of *s*, depending the vacancy *v*. When the roadblocks are randomly positioned, there is a finite probability, $ve^{1-1/v}$, that an entire binding site is open. When this is the case, the distance, *s*, to the nearest roadblock is exponentially distributed with an average of $dv/(1-v)$.

**Fig. 3.** (A) The effective target size, $\langle L(v) \rangle$, as a function of vacancy. We assume that all roadblocks and TFs have an average footprint on DNA of *d*=20bp. The sliding length $l_{sl} = 2\sqrt{D_1/\Lambda}$ is estimated with the 1D diffusion constant measured *in vitro* for the *lac* repressor (0.046μm²/s from ref. [13]) and the diffusion-limited macroscopic dissociation rate constant from Eq. (5). In estimating $K_{RD}$ in Eq. (5), we assume that *v*=0.7 for the *in vivo* experiment measuring the effective diffusion constant [13, 17] and use Eq. (13) with $c_{ns}$=4.6x10⁶/1μm³. This results in $K_{RD} \approx 1900 M^{-1}$ and $\Lambda \approx 750 s^{-1}$. Notice that these numbers are different from what was estimated before without considering the limited DNA vacancy. *blue line* static road-blocks, *purple line* dynamic roadblocks (B) The search time, defined for one TF to find one target site, as a function of vacancy. The search time is calculated from the association rate constant assuming only one TF in a volume of 1μm³. *Blue line:* static road-blocks *Purple line:* dynamic roadblocks

**Fig. 4.** Facilitation by sliding along nonspecific DNA. The search time as a function of vacancy is plotted for the sliding model (*blue line* static road-blocks *purple line* dynamic roadblocks) and for hypothetical cases with no sliding (*yellow line*) or no nonspecific binding (*green line*). Sliding along DNA still results in shorter search time despite the presence of roadblocks. Parameter values used are the same as in Figure 3. *Inset* The rate enhancement for facilitated diffusion compared to only cytoplasmic diffusion (no nonspecific binding) is plotted at various values of nonspecific binding constant (*solid*: $K_{RD} = 2000 M^{-1}$, *dash*: $K_{RD} = 20000 M^{-1}$, and *dot*: $K_{RD} = 200 M^{-1}$). Only stationary roadblocks are considered here for presentation. The calculated rate enhancement is based on the measured *in vitro* value for $D_1$, and it would be much smaller if the sliding rate is substantially reduced *in vivo*.

**Fig. 5.** The relative search time as a function of total number of DNA-binding proteins. In the *E. coli* genome, we consider a single site controlled by a specific DNA-binding protein whose concentration is at a fixed fraction (one per one thousand) of all DNA-binding proteins. The amount of time it takes for the site to be occupied, normalized to the lowest point in the graph, is plotted against the total number of all DNA-binding proteins per chromosome equivalent. When the amount of DNA-binding proteins is small (<10⁴), increasing the overall number would speed up the search process as expected. However, for the total number close to that measured in *E. coli* (*green bar*), further increasing the copy number has marginal effect in reducing the reaction time. The DNA-binding protein is modeled based on the *lac* repressor, with the same parameter values as in Figure 3, and the exact shape of the curve will vary with the specific properties of individual proteins.

**Fig. 6.** (A) *The effects of DNA looping on the search kinetics.* Association to binding sites (*red*) that are farther apart than the average sliding distance $l_{sl}$ will occur independently (*bottom*) and the rate for the first association to one of the two sites is $2k_a$. When the sites are closer than the average sliding distance (*top*), association to the sites are strongly correlated and they appear as only one target for which the association rate constant is $k_a$.

(B) *Forming and breaking the loop. Top* The TF can find and dissociate from the first binding site by sliding over nonspecific DNA sequences following a helical path. When bound at one site, a DNA loop only can be formed by direct association to the other site. Association by sliding is in this case hindered since it would require diffusion along a helical path which requires both twisting and lateral diffusion of the linker DNA. *Bottom* Breaking the loop requires that the TF overcomes both the specific hydrogen bonds and the non-specific electrostatic interaction with the backbone essentially in one step, since sliding out from the looped complex is geometrically hindered.

Figure 1

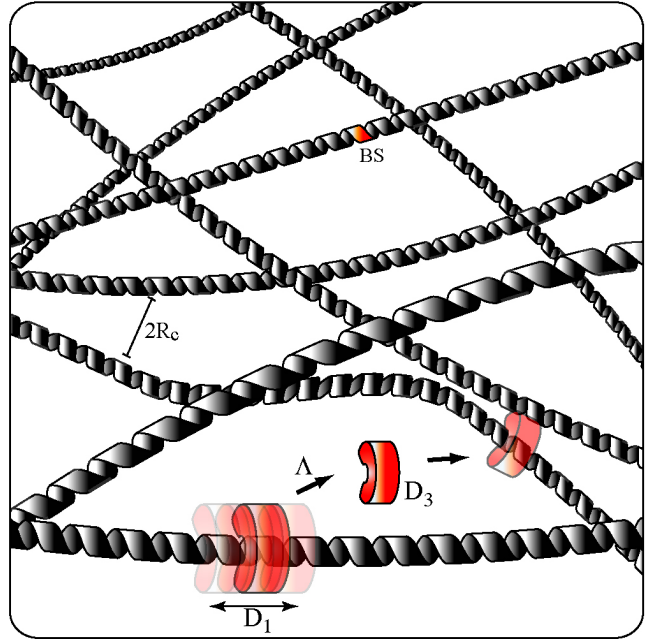

Figure 2

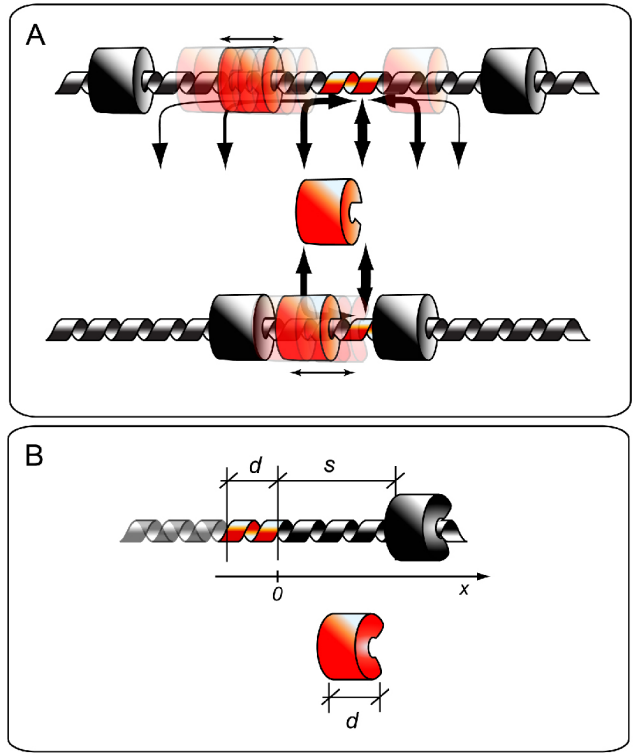

Figure 3

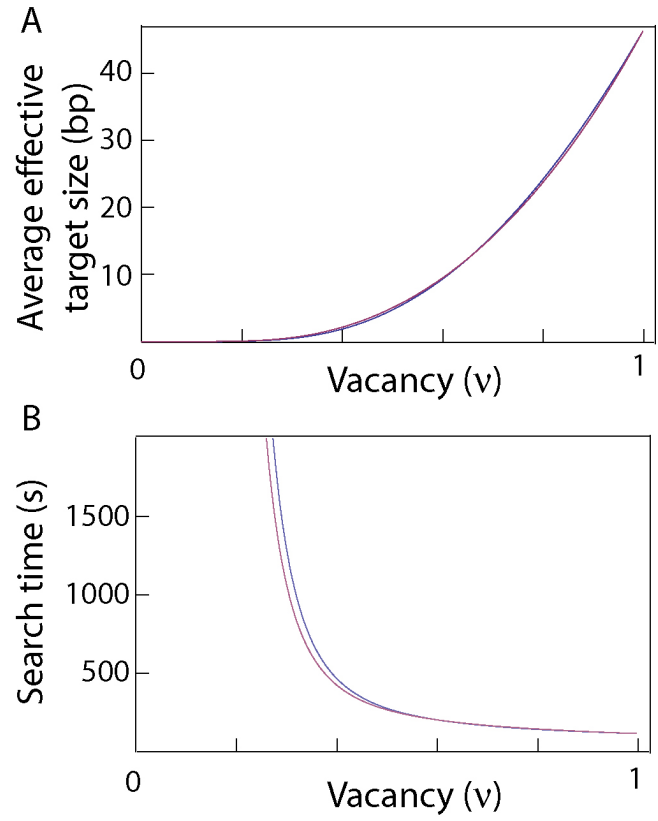

Figure 4

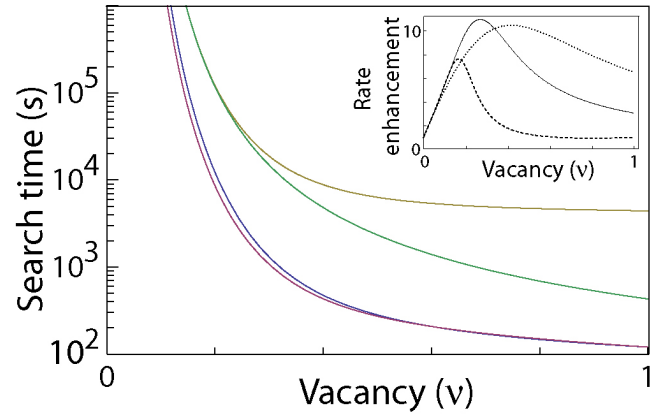

Figure 5

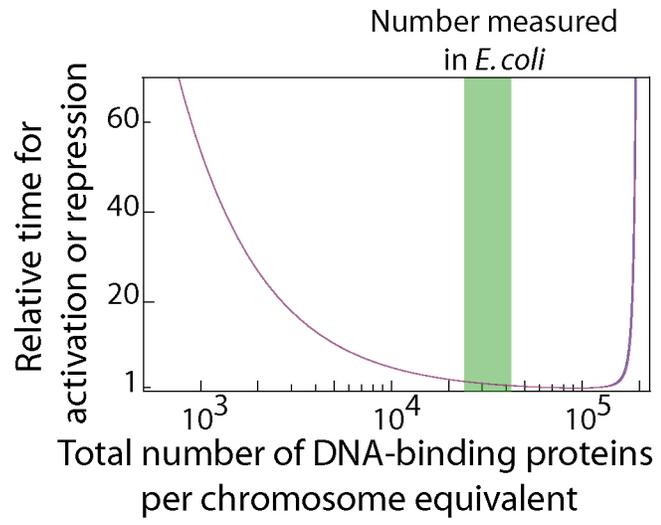

Figure 6

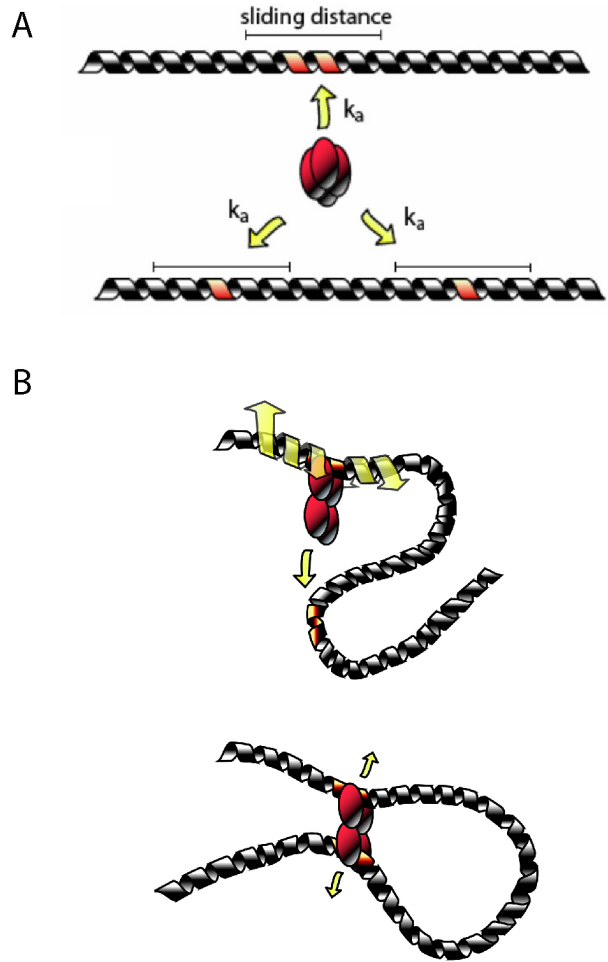

# Supporting information
## Gap size distribution of randomly positioned roadblocks

In the seminal work by McGhee and von Hippel (1), the statistical distribution of non-interacting ligands on a discrete DNA lattice was analyzed. In the current notations, the probability for a gap being of length $s$ basepairs is

$$p_{gap}(s) = \left(1 + \frac{dv}{1-v}\right)^{-1}\left(1 + \frac{1-v}{dv}\right)^{-s}.$$

Here we consider the case when the average gap size is greater than one basepair ($v>0.05$ for $d=20$ bp). In this limit, we can take the continuous approximation of the DNA lattice. The gap size distribution becomes

$$p_{gap}(s) \approx \frac{1-v}{dv}\exp\left(-\frac{s}{dv/(1-v)}\right).$$

The exponential distribution indicates that the probability of ending the gap at a given position is independent of how long the roadblock-free region is prior to that position. Therefore, this distribution is the same no matter where we start measuring the length, as long as the starting position is in a gap.

## Effective target size in dynamic roadblocks

To analyze the scenario with diffusing roadblocks, we use a different approach from the stationary scenario. The effective target size can be expressed a product of the probability to land on DNA (binding to a roadblock-free nonspecific site) upon encounter and the length of DNA along which the protein can diffuse into the target without dissociation. As in the stationary roadblock case, the probability for a free segment longer than the protein footprint ($d$) at a particular position is $ve^{1-\frac{1}{v}}$. This is also the probability that the protein can bind to DNA when diffusing from cytoplasm.

To calculate the average length of DNA from which the protein can slide into the target before dissociation, we use a first-passage time based method introduced recently. For a protein initially bound at position $x$ on nonspecific DNA, the probability to reach the target is the time integral of the first-passage time density, $f(x,t)$, multiplied by the likelihood of staying bound nonspecifically at a given time. The average DNA length is thus an integral along the DNA, weighted by this probability reaching the target. If the nonspecific residence time is exponentially distributed with average $t_R$, the effective target size is (including the factor from the previous paragraph):

$$<L(v)>_d = ve^{1-\frac{1}{v}}\int_{-\infty}^{\infty}\left(\int_0^{\infty} f(x,t)e^{-t/t_R}dt\right)dx$$

Since the free propagator in Single-File diffusion is Gaussian-shaped with variance of $\frac{2D_1 t}{1+(t/t_x)^{1/2}}$, we can obtain the first passage time density with absorbing boundary condition at $x=0$, using the method of images.

$$f(x,t) = -\frac{\partial}{\partial t}\int_0^{\infty} G(x',t;x) - G(x',t;-x)dx',$$

where $G(x', t; x)$ is the free propagator from $(x, 0)$ to $(x', t)$.

$$G(x',t;x) = \sqrt{\frac{1+(t/t_x)^{1/2}}{4D_1 t}}\exp\left(-(x'-x)^2\frac{1+(t/t_x)^{1/2}}{4D_1 t}\right)$$

From the three equations above, we obtain a complicated expression for the effective target size:

$$<L(v)>_d = v e^{1-\frac{1}{v}} \frac{\sqrt{D_1 t_R}}{\sqrt{\pi}} \begin{bmatrix} \left(\frac{t_x}{t_R}\right)^{\frac{1}{4}} \Gamma\left(\frac{1}{4}\right) {}_1F_1\left(\frac{5}{4}, \frac{1}{2}, -\frac{t_x}{t_R}\right) \\ + 6\left(\frac{t_x}{t_R}\right)^{\frac{3}{4}} \Gamma\left(\frac{3}{4}\right) {}_1F_1\left(\frac{7}{4}, \frac{3}{2}, -\frac{t_x}{t_R}\right) \\ - 8\left(\frac{t_x}{t_R}\right)^{\frac{3}{4}} \Gamma\left(\frac{3}{4}\right) {}_1F_1\left(\frac{3}{4}, \frac{1}{2}, -\frac{t_x}{t_R}\right) \\ + 4\left(\frac{t_x}{t_R}\right)^{\frac{5}{4}} \Gamma\left(\frac{1}{4}\right) {}_1F_1\left(\frac{5}{4}, \frac{3}{2}, -\frac{t_x}{t_R}\right) \\ + 8\left(\frac{t_x}{t_R}\right)^{\frac{1}{2}} {}_2F_2\left(\frac{1}{2}, 1; \frac{1}{4}, \frac{3}{4}; -\frac{t_x}{t_R}\right) \\ - 8\left(\frac{t_x}{t_R}\right)^{\frac{1}{2}} {}_2F_2\left(1, \frac{3}{2}; \frac{3}{4}, \frac{5}{4}; -\frac{t_x}{t_R}\right) \end{bmatrix},$$

where ${}_pF_q$ is the generalized hypergeometric function and $\Gamma$ is the gamma function. This expression can be approximated as

$$<L(v)>_d = v e^{1-\frac{1}{v}} \left[ 2\sqrt{\frac{D_1 t_R}{1+(t/t_x)^{1/2}}} \right].$$

We plot the exact solution and the approximated function in Figure S1. The approximation holds from six decades in $t_x$.

**References**
1. McGhee JD & von Hippel PH (1974) *Journal of Molecular Biology* **86,** 469-489.

**Figure Legends**
The approximation for the effective target size with dynamic roadblocks. Here we plot $\frac{<L(v)>_d}{2\sqrt{D_1 t_R} v e^{1-\frac{1}{v}}}$ as a function of $\frac{t_x}{t_R}$ for both the exact solution and the approximation. The difference is within one percent throughout six decades. *Blue: exact solution, Purple: approximation*